\documentclass[5p]{elsarticle}
\usepackage{amssymb}
\usepackage{amsmath}
\usepackage{graphicx}
\usepackage{epstopdf}
\usepackage{ulem}
\usepackage{xcolor}
\begin{document}
\title{Atom beam triangulation of organic layers at 100 meV normal energy: self-assembled perylene on Ag(110) at room temperature}

\author[ismo,im2np]{Nataliya Kalashnyk}
%\author[ismo]{Hocine Khemliche}
\author[ismo]{Philippe Roncin\corref{cor1}}
\ead{philippe.roncin@u-psud.fr}
\cortext[cor1]{Corresponding author}
\address{Institut des Sciences Mol\'{e}culaires d'Orsay (ISMO), CNRS, Univ. Paris-Sud, Universit\'{e} Paris-Saclay, F-91405 Orsay, France}
\address[im2np]{Institut Mat\'{e}riaux Micro\'{e}lectronique Nanosciences de Provence (IM2NP), CNRS, Universit\'{e} d'Aix-Marseille, F-13397 Marseille, France}

\begin{abstract}
	
The controlled growth of organic layers on surfaces is still waiting for an in-situ reliable technique that would allow their quality to be monitored and improved. Here we show that the growth of a perylene monolayer deposited on Ag(110) at room temperature can be tracked with low energy atoms in a regime where the energy perpendicular to the layer is less than 0.1 eV and below the organic film damage threshold. The  image processing required for this atom triangulation technique is described in detail.
\end{abstract}

\maketitle

\section{Introduction} 
Thin films of polycyclic aromatic hydrocarbons such as perylene (C$_{20}$H$_{12}$) are investigated intensely for their integration in a broad range of applications, for instance in organic light emitting diodes (OLED). A large number of studies over the last decade have demonstrated that molecular organization is a crucial factor for the physical properties of organic layers. Thin film assembly can be investigated locally in great detail by STM \cite{Bobrov} and/or AFM \cite{Witte} but not during the growth of the layer. Other techniques such as reflection high energy electron diffraction (RHEED) widely used in molecular beam epitaxy, are used to probe growth in-situ but the high beam energy destroys the fragile organic layers. In fact, there is a clear lack of techniques able to monitor {\textit{in-situ} and in real time the structure of the molecular network and its quality. Using the same grazing incidence geometry as RHEED, i.e. compatible with in situ operation, two techniques sensitive only to the topmost layer have emerged recently. These are grazing incidence fast atom diffraction (GIFAD or FAD) \cite{Zugarramurdi_2015,Seifert_alaninePRL} and fast ion or atom beam triangulation (IBT) \cite{Seifert_alaninePRL,Seifert_alaninePRB,Pfandzelter,Seifert2012,Bernhard}. The first one is very demanding in terms of long range order while the second one, developed for inorganic crystals operates at comparatively high interaction energy with the surface. In this paper, a new regime of extremely weak interaction with the surface is briefly presented, and applied to the investigation of perylene assembly on Ag(110). A simple image processing method providing a more efficient triangulation technique is described in detail. The convergence between GIFAD and IBT is discussed at the end.

\subsection{IBT and GIFAD} 

IBT is a technique allowing the identification of the low index directions of the topmost layer of a crystalline surface. The working principle of this method is that an atom or an ion impacting on a surface at a grazing incidence angle will experience a specific physical condition known as axial channeling if it is aligned with a low index direction. Several physical properties can be exploited depending on the nature and energy $E_0$ of the projectile. Protons with  $E_0$=25 keV at an incidence angle $\theta_{in}$=1.6  $^{\circ}$  with respect to the Cu(001) surface have a normal energy $E_{\perp}=E_0 sin(\theta_{in})^2\approx 20 eV$. Monitoring the electric current flowing to the target during the azimuthal scan is a simple technique giving an indirect measure of the emission of secondary electrons \cite{Pfandzelter}.
In this energy range and along the surface channeling direction, the projectiles undergo multiple bounces \cite{Michely2006} on the atomic rows (the walls of the channel) and have more chance to trigger electron emission.
 This interpretation was confirmed using an electron detector in front of the surface, which provided a higher sensitivity allowing the beam intensity to be reduced \cite{Bernhard}. 
At such comparatively high projectile normal energy, the initial projectile charge state has little consequence and similar contrast was observed with neutral hydrogen atoms \cite{Seifert2012}. Only recently has the same triangulation strategy been applied to lower energy atoms of 2 keV with a perpendicular energy $E_{\perp}$ between 0.6 and 1 eV \cite{Seifert_alaninePRL,Seifert_alaninePRB}. In this case, there is no electron emission and the azimuthal scan is performed by monitoring the variation of the intensity of the scattering profile in the specular region. The azimuthal contrast is generated by the fact that along the channeling directions, successive momentum transfers are cumulative instead of being randomly distributed. As a result the scattered beam has a lower, broader peak intensity with dips generated at each low index direction. 
In this range where the perpendicular energy is only few meV, the distance between the projectile to the surface layer is around 1-3 \AA{} from the top layer. This is true only if the deposited layer is already well organized or dense enough, so that mutual shadowing can occur via successive deflections which prevent the projectile from approaching too close to the molecules. On the other hand, if an isolated molecule or a terrace edge is encountered on the bare metal surface, then the interaction energy is probably large enough to destroy the molecule or to penetrate the terrace.
Here, we explore the same technique at even lower normal energies, with 300 eV helium atoms having an $E_{\perp}$ in the 50-100 meV range where the minimum distance to the surface is above 3-4 \AA{} so that even an isolated molecule will not experience a close collision, provided that it is lying flat on the surface. The helium atom will fly over the molecule, or the terrace edge, without penetration and with limited momentum transfer.
Initially, this energy interval  was chosen because it corresponds to the conditions where diffraction was observed in GIFAD. If the transverse coherence $\Delta_x$ of the atomic beam is larger than the lattice parameter $d$, diffraction can be observed as Bragg peaks separated by a reciprocal lattice vector $G=2\pi/d$ in the scattering profile. The transverse coherence is a simple statistical property of the beam expressed as $\Delta_x = 2\pi/\Delta_k$ with $\Delta_k=k_0 \Delta\psi$, where $k_0$ is the momentum of the atoms, and $\Delta\psi$ is the angular divergence of the beam. The additional information provided by the lattice parameter is extremely valuable to complement the triangulation. Indeed, triangulation provides the directions of alignment but does not specify either the unit cell size or the number of molecules in the cell whereas even weak diffraction is sufficient to obtain these parameters with a quite high precision. In the case of a surface exhibiting sharp distinct diffraction spots, the profile of the surface electronic density can be retrieved with exceptional accuracy \cite{DebiossacPRB,schuller_2010}. However, it should be noted that this high resolution regime has not been achieved so far for molecular layers. Clear diffraction lines have recently been reported for comparatively small, non-planar alanine molecules chemically adsorbed on a Cu(110) surface \cite{Seifert_alaninePRL,Seifert_alaninePRB}. The present paper reports the experimental investigation of perylene growth on Ag(110). This large planar molecule shows significantly lower interaction with the surface leading to a partial flexibility in the assembly of such an organic monolayer\cite{Bobrov}. Only weak diffraction is observed with a contrast weakening over time. This is probably due to the metastability of the layer. To complement and prepare the diffraction measurement, we have developed a new simple and robust procedure for triangulation with low energy atoms.

\section{Experimental setup}

\begin{figure}%[ht]
	\includegraphics[width=0.9\linewidth]{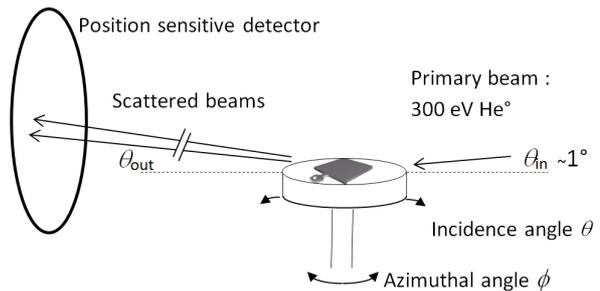}
	%\begin{center}
	%	\includegraphics [width=70mm]{schematic.eps}
	%\end{center}
	\caption{{Scheme of the experimental setup. The surface is placed on a Omicron plate depicted in gray in the center and exposed to the 300 eV He$^{\circ}$ beam at grazing incidence. The scattered beams are detected almost a meter downstream on a position sensitive detector imaging the outgoing angles. The primary beam and the detector remain fixed during the experiment.}}
	\label{Schematic}
\end{figure}
The deposition of a perylene monolayer on the Ag(110) surface was conducted in an ultra-high vacuum (UHV) system equipped with standard facilities for sample preparation and analysis as well as a separate load-lock chamber compatible with a high performance portable UHV suitcase. The Ag(110) surface was cleaned by several cycles of Ar ion sputtering followed by annealing. The perylene molecules were evaporated on the Ag(110) surface held at room temperature, and the molecular structure was subsequently checked with a microchannel plate-low energy electron diffraction (MCP-LEED) system operated at an electron energy of 18 eV, which is surface sensitive and limits the damage to the organic layer.
The sample was then transferred by means of the portable UHV suitcase into the independent GIFAD chamber without breaking the UHV conditions.  The surface is mounted perpendicular to the rotation axis of a standard X,Y,Z,$\phi$ sample manipulator itself mounted on a custom designed tilt system controlling the angle of incidence $\theta$ (see Fig. \ref{Schematic}). The atomic beam of 300 eV neutral helium passes into the vacuum chamber through a 100 $\mu m$ circular hole. Prior to inserting the sample, the beam position is recorded onto the position sensitive MCP detector in front of a phosphor screen and filmed by a CCD camera.  Then the surface is progressively inserted in the beam and the angle of incidence is tuned to 0.7$^{\circ}$. Most often, the surface is only partially inserted into the beam leaving a small part of the neutral helium atoms flying over the surface without interaction to produce a small reference spot on the detector as can be seen at the bottom of the scattering patterns in Fig.  \ref{raw_profiles}.\\

\section{Results}
A diffraction signal was observed after the sample was introduced into the beam and vanished few hours later under exactly the same conditions. Therefore, a systematic azimuthal scan was performed. In principle, such a scan requires only a continuous or stepwise rotation synchronized with a movie or with a sequence of images. This was not possible because the surface plane defined by its normal vector was found to be misaligned with the rotation axis by approximately 1$^{\circ}$, as can be seen in the azimuthal scans Fig. \ref{swing} below. An automatic compensation could have been programmed as explained in \cite{sereno}. In the present case, approximate manual corrections of the angle of incidence $\theta$ were needed every few images to maintain the scattering profile in the same region on the detector. Each image of the scattering profile is composed of fifty exposures of 2 s accumulated directly  inside the CCD camera before transfer to the PC. Consequently, in order to perform the azimuthal scan in a reasonable time, comparatively large angular $\phi$ steps of two degrees were chosen. In such a long exposure, the camera produces a significant noise, so a dark image was also recorded under exactly the same exposure conditions but with the beam valve closed. This dark image is automatically subtracted from all images.

\subsection{Data analysis}
\begin{figure}[ht]
	\begin{center} \includegraphics [width=80mm]{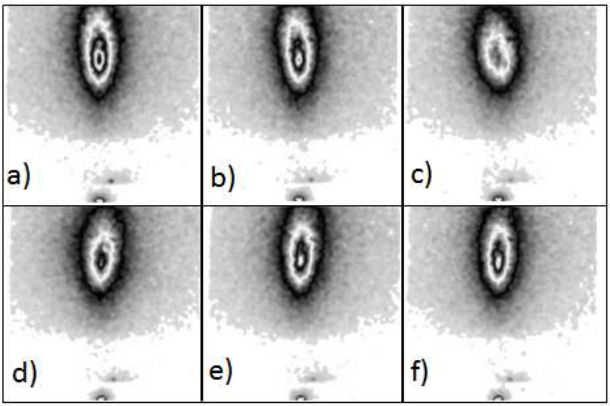} \end{center}
	\caption{{Raw images of the scattering patterns recorded every two degrees azimuth for a perylene monolayer on Ag(110) substrate. The relative orientation is as sketched in Fig.  \ref{Schematic}, i.e. the surface plane is horizontal while the outgoing scattering angle $\theta_{out}$ increases along the vertical direction. The color scale is a fixed black and white banded palette highlighting a quasi-elliptic white contour. The corresponding labels (a) to (f) are reported on the azimuthal scan in Fig. \ref{Azimuth}.}}
	\label{raw_profiles}
\end{figure}
Six successive images of a region of interest in the  detector are displayed in Fig. \ref{raw_profiles}. At first glance these images look rather similar having a smooth scattering profile without any evidence of Bragg peaks. A closer inspection reveals that the pattern experiences profound variations. The white quasi-elliptic contour line undergoes an anti-clockwise rotation between the images in Fig. \ref{raw_profiles}(a) and Fig. \ref{raw_profiles}(c) whereas the evolution between the images in Fig. \ref{raw_profiles}(d) and Fig. \ref{raw_profiles}(f) is clockwise. Between Fig. \ref{raw_profiles}(c) and Fig. \ref{raw_profiles}(d), the variation is more drastic with a sudden mirror change of the orientation. This behavior is characteristic of surface channeling of ions or atoms along a low index direction. Since the energy is more than 100 times less than in \cite{Pfandzelter,Bernhard} we try to clarify the specific dynamics occurring as the projectile direction approaches a low index direction as a function energy.

At high normal energy $E_{\perp} \gtrsim 10 $ eV the projectiles penetrate deeply the surface electronic density. The curves of iso-electronic density no longer form a 2D manifold over the atomic layer but split into separated spheroids centered around the atoms so that some projectile trajectories may even penetrate below the surface. However, most of the projectiles simply bounce in-between the rows of surface atoms \cite{Danailov} and undergo multiple collisions from one side of the channel to the other in a genuine guiding that generates specific rainbows \cite{Borka}. 

At $E_{\perp} \lesssim 100 $ meV used in this work, the evolution as the beam direction approaches a low index direction has been measured in detail in the diffraction regime \cite{Seifert_2011,Zugarramurdi2013b} and is well described by quantum mechanics \cite{Zugarramurdi2013a} or semi-classical models for weak surface corrugation \cite{DebiossacPRA}. In this regime the net deflection of the beam is very limited and is localized at scattering angles significantly larger than the specular angle \cite{Seifert_2011,Zugarramurdi2013b,Zugarramurdi2013a}. In the present experiment the later effect is likely to be responsible for the clear rotation of the top of the contour ellipses in Fig. \ref{raw_profiles}. As noted in \cite{Zugarramurdi2013b} this indicates that the twist angle of the ellipse is a differential signal which has a zero twist when the beam is perfectly aligned with a low index or along a random directions. To identify the low index directions, we concentrate on robust features of the scattering pattern such as the horizontal or vertical intensity profiles. The associated widths of the observed distribution seem an obvious choice, however,  we have to take into account that the angle of incidence $\theta_{in}$ could not be kept constant due to the tilt angle with the manipulator axis. We decided to transform our 2D images into a coordinate system which is less sensitive to the variation of the angle of incidence.
\subsubsection{Polar coordinates}
 The Cartesian coordinates on the raw images in 
 Fig. \ref{raw_profiles} correspond to scattering angles parallel and perpendicular to the surface. For scattering problems in general, the magnitude of the deflection and the associated polar profile are the most relevant. Examples include (i) scattering rainbows observed at at fixed polar scattering angle $\alpha$ (see 
 Fig. \ref{polar_like}) over a wide range of incidence angles and (ii) supernumerary rainbows \cite{Schuller2008}. 
 Polar coordinates are still valid in the quantum scattering regime where diffraction charts show clear nodal structures well-aligned with the polar angle. In fact these nodal structures transform into attenuating supernumerary rainbows when the diffraction is no longer visible. The use of these coordinates should provide image properties that are much less sensitive to the actual angle of incidence.

\subsubsection{Polar-like 2D transformation}

A pragmatic choice is to consider the position of the direct beam as a reference in the raw image as well as for the transformation. Taking $(0,0)$ for the beam coordinates, each pixel of coordinates $(k_x,k_y)$ is part of a circle having a center $(k_{xc},k_{yc})$ given by $k_{xc}=0$, $k_{yc}=k_y/2+k_x^2/k_y$, a radius $k_{eff} = k_{yc}$ and $\alpha=sin^{-1}(k_x/k_{eff})$. This polar-like transform $(k_x,k_y)\rightarrow(\alpha,2k_{eff})$ is defined everywhere above the direct beam $(k_y>0)$. In order to provide a measure close to the original one, the polar angles are reported between -90  $^{\circ}$ and +90  $^{\circ}$ only, corresponding to the upper parts of the circles in Fig. \ref{polar_like}. Note that this transformation does not affect at all the vertical line above the direct beam, i.e. the scattering profile in the incidence plane. Also, it does not seriously impact the general shape of the scattering patterns as seen in Fig. \ref{polar_like}.
\begin{figure}[ht] %ht
	\begin{center} \includegraphics [width=70mm]{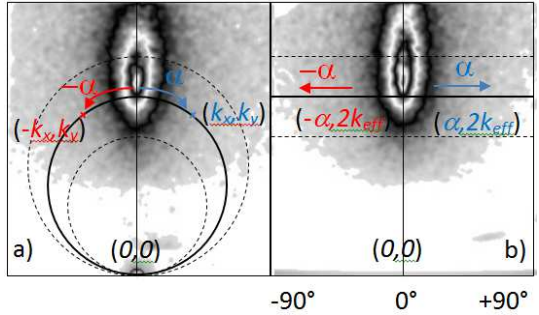} \end{center}
	\caption{{Sketch of the polar-like transformation associating (a) a coordinate $(k_x,k_y$) to (b) new coordinates $(\alpha,2k_{eff})$ defined by: the radius of the circle $k_{eff}$, the point $(k_x,k_y)$, its left/right symmetry $(-k_x,k_y)$ , and the direct beam. This forces the center to lie on the vertical axis above the primary beam, i.e. in the plane of incidence. Finally,  $\alpha$ is the polar angle along this circle. In this transformation, the incidence plane (vertical line) is strictly invariant  and most of the image is hardly affected.}}
	\label{polar_like}
\end{figure}

\subsubsection{1D scattering and polar profiles}

The scattering profile in the plane of incidence is comparatively broad, and it is reasonably well fitted by a log-normal distribution commonly used in atom surface scattering \cite{Villette2000,Manson2008}. This profile is used here only to measure the position of the maximum. All polar plots $P(\alpha)$ exploited for the azimuthal scan correspond to a slice of -0.2$^{\circ}$ below the maximum of this scattering distribution. Two examples of the polar angle distribution are displayed in Fig. \ref{polar_plots}, and are empirically fitted by a Lorentzian profile to extract three parameters: the mean value $\langle \alpha \rangle$, the width $W_\alpha$, and the intensity.

\begin{figure}[ht]
	\begin{center} \includegraphics [width=80mm]{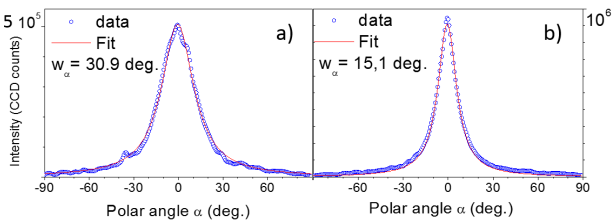} \end{center}
	\caption{{Polar scattering profiles P($\alpha$). The left panel (a) corresponds to the images displayed in Fig. \ref{raw_profiles}(c), while the right panel (b) is derived from Fig. \ref{raw_profiles}(f), shown also in Fig. \ref{polar_like}(a).}}
	\label{polar_plots}
\end{figure}

\subsection{Azimuthal scans}
Fig. \ref{Azimuth} displays a detailed azimuthal scan of the measured Lorentzian width $W_{\alpha}$ evidencing the presence of low index directions. Note that, instead  of using polar coordinates, almost the same result could be obtained by dividing the horizontal width by the mean scattering angle directly from raw images. The more rigorous, polar approach was adopted here with the advantage that the tilt angle can be measured directly.
\begin{figure}[ht]
	\begin{center} \includegraphics [width=80mm]{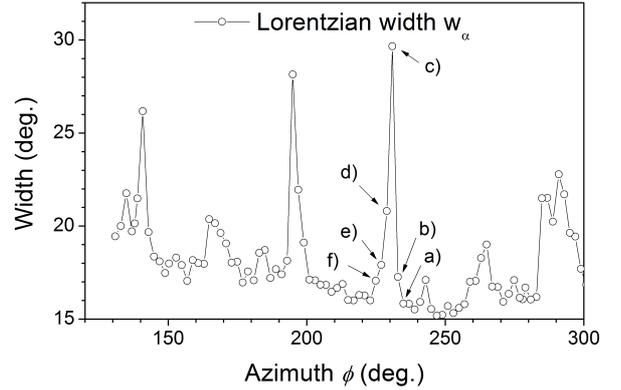}\end{center}
	\caption{{Azimuthal scan of the polar width as measured by a Lorentzian fit (see e.g. Fig. \ref{polar_plots}). The scan outlines sharp variations taking place within one or two degrees. The six labels (a) to (f) refer to the images in Fig. \ref{raw_profiles}. }}
	\label{Azimuth}
\end{figure}
Fig. \ref{swing} shows the smooth evolution of the mean polar angle $\langle \alpha \rangle _{bottom}$ measured by the fit on the bottom part of the scattering distribution that amounts to 0.2$^{\circ}$ below the maximum determined from the log-normal fit of the vertical profile. This dependence is adjusted with a pure sine function with a full amplitude of 4.3$^{\circ}$.  
\begin{figure}[ht]
	\begin{center} \includegraphics [width=80mm]{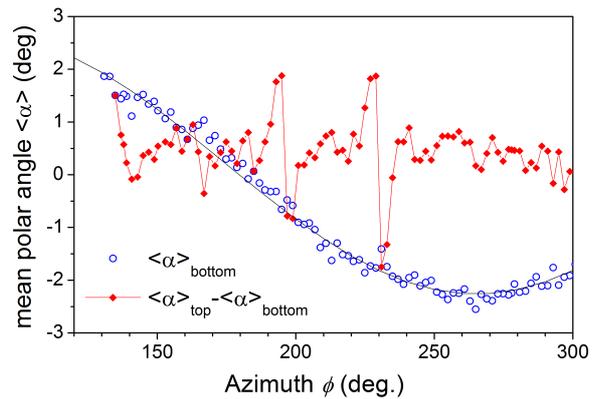} \end{center}
	\caption{{Azimuthal scan of the center $\langle \alpha \rangle _{bottom}$  of the polar profile ($\circ$) that is measured by the Lorentzian fit in Fig.  \ref{Azimuth} in the bottom part of the images. The diamonds  ($\blacklozenge$) plot the difference with the same quantity $\langle \alpha \rangle _{top}$ measured in the upper part of the same scattering images.}}
	\label{swing}
\end{figure}
This corresponds to a tilt angle of $\tau=1.1^{\circ}$  as sketched in Fig. \ref{tilt}.
It should be mentioned that in the natural angular frame of the detector (referred to as the Cartesian frame above) this 4.3$^{\circ}$ amplitude of the angular swing is only 0.05$^{\circ}$  since the typical value of $\theta_{eff}$ representing the radius of the circle in Fig. \ref{polar_like} corresponds to 0.7$^{\circ}$.
 \begin{figure}[ht]
 	\begin{center} \includegraphics [width=50mm]{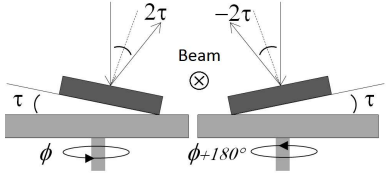} \end{center}
 	\caption{{Sketch of the maximum effect of a tilt angle $\tau$ between the surface (dark gray) and the rotation axis (light grey) on the specular reflection in the polar coordinates. Note that the main velocity component of the beam is not visible. }}
 	\label{tilt}
 \end{figure} 
As a naive attempt to capture the rotation of the ellipse visible in Fig. \ref{raw_profiles} we have also measured the mean polar angle $\langle \alpha \rangle _{top}$, a 0.2$^{\circ}$ wide slice on top of all images, and subtracted it from the earlier defined $\langle \alpha \rangle _{bottom}$ , i.e. the 0.2$^{\circ}$ slice below the maximum. Fig. \ref{swing} shows bipolar oscillation around the low index directions identified in Fig. \ref{Azimuth}. This nicely illustrates the rotation effect  of the quasi-ellipse. 

Combining the azimuthal scan in Fig. \ref{Azimuth} together with the initial GIFAD diffraction and LEED data is comparatively easy because each piece of information is complementary, however, this will not be discussed here. The reason is that we lack the redundancy of a complete azimuthal scan and, more importantly, the rapid degradation of the sample in a single day at a pressure below $10^{-10}$ mbar. The latter incentive demonstrates that the molecular organization is metastable reducing our confidence that the diffraction and triangulation measurements were actually observing the same structure. This doubt was also triggered by the fact that our observation does not directly match the direct STM imaging after a comparable preparation. This is probably due to the fact that the system seems very sensitive to the density of deposited molecules so that there could be a gradual transition from a densely packed comparatively well-organized layer to a looser organization \cite{Bobrov}.

\section{Discussion}

The present experimental conditions were chosen to track diffraction features. This has led to meticulous recording of more than 100 successive images each exposed for 100 seconds with additional manual control, stretching over almost ten hours. One of the reasons for that was that the beam intensity was low, the other being due to the corrections imposed by the tilt angle. However, even without any tilt of the surface, i.e. without any need for manual control, this would have required three hours of beam time. The experiment duration can be drastically reduced by noting that the smaller structure observed in the images is of the order of 0.08$^{\circ}$ full width at half maximum (FWHM) so that, if triangulation only had been targeted, the two 100$\mu m$ diaphragms reducing the beam divergence below 0.02$^{\circ}$ could have been replaced by two 400$\mu m$ diaphragms giving an intensity 256 times larger. Indeed resolution scales with the diameter of the diaphragm but the beam intensity scales with the fourth power of the diameter. A scan would take less than 10 min without any loss. It is possible to gain another order of magnitude to bring the acquisition time close to a minute by compromising the resolution or by using a more sophisticated image processing algorithm. Scanning times shorter than a minute are probably not relevant since fast mechanical action under vacuum is usually not recommended. These advances are still compatible with in-situ analysis given the  benefit that molecular growth is usually more uniform on a rotating sample smearing possible beam inhomogeneity. Finally, the comparatively low angular resolution of the present study can be improved by a continuous recording while the image statistics can be recovered by averaging successive images.

The width method detailed here is similar in essence to that reported in \cite{Seifert_alaninePRB} where the peak intensity of the scattering profile was used. Both approaches can probably be further improved and generalized by a generic multipolar expansion of the recorded images to extract the main scattering parameters: mean values, width, peak intensity, and more refined ones like the $k_x.k_y$ correlation terms quantifying the orientation of the scattering pattern. 
The other aspect illustrated here is that triangulation is less demanding than diffraction. This can be extremely useful to extend the range of GIFAD operation. Such a tolerance to local disorder can be understood in terms of the  surface coherence length $L_S$  defined as the mean distance between defect. 
This distance should be compared with the mean length $L_T$ of the atom trajectory on the surface estimated from classical trajectories or simple models \cite{RousseauNIMB2007}. Elastic diffraction can only occur if $L_S \gg L_T$ and this has not been observed so far for molecular layers. 
If $L_S \gtrsim L_T$ inelastic diffraction can be observed providing accurate lattice parameter, but its quantitative interpretation in terms of surface electronic density might be difficult. If $L_S \lesssim L_T$ then no diffraction should be observed, however the momentum transferred to the projectile atom along its trajectory is still very much influenced by local molecular alignment as probed in triangulation. In other words triangulation is likely to identify the local order over a much broader range of surface coherence length (i.e. global order). Meanwhile inelastic diffraction and possibly elastic diffraction will progressively show up only as the quality of organic layer improves.
Both IBT and GIFAD have demonstrated a sensitivity to the topmost layer and are able to count the exact number of successive layers \cite{Atkinson,Igel} during growth.  Both approaches are fully compatible and obviously needed to help monitoring in-situ the growth of very well organized molecular layers.
The very encouraging new aspect demonstrated here is that azimuthal scans are well-contrasted even with the low perpendicular energies needed for GIFAD. These low values, $E_{\perp}\lesssim 0.1 $ eV should not induce any damage to the fragile organic film. At first sight, it seems rather illusory to record an azimuthal dependence at one degree of incidence of a surface itself tilted by more than a degree! The straightforward image processing detailed here shows that an azimuthal scan with significant intensity variations can be retrieved.

\section{Conclusion}
Using atomic beams of very low effective interaction energy with the surface in a regime where the damage to fragile molecular layers is expected to vanish constitutes a powerful method to probe the self-assembly of adsorbed monolayers. A simple processing of the scattering profile generates highly contrasted azimuthal scans revealing the directions of molecular alignment. In this respect, implementing triangulation with a position-sensitive detector is certainly a good choice paving the way to atomic diffraction if a molecular layer with long-range order is formed. Additionally, triangulation becomes compatible with GIFAD offering fewer restrictions on the level of organization so that molecular layer growth can be monitored over a very broad range of relative disorder.

\section{Acknowledgement}
We are  grateful to A. Mayne for careful reading of this manuscript and to H. Khemliche, A. Momeni and L. Guillemot for their assistance in running the GIFAD setup and the transfer chamber.

%\lfoot{\small{\textsf{*Corresponding author e-mail address: philippe.roncin@u-psud.fr}}\\ \vspace{-0.5cm}\hrule}

\end{document}